\documentstyle[12pt,epsfig]{article}
\title{{\LARGE\bf Some problems in plasma suppression of beam-beam
interactions at muon colliders.}
  \thanks{Talk at the Workshop Studies on Colliders and
    Collider Physics at the Highest Energies: Muon Colliders at 10 TeV
    to 100 TeV, 27 September - 1 October, 1999 Montauk, New York, USA,
    be published by the American Institute of Physics.  }  }
\author{Valery Telnov, \\
  {\small\it Budker Institute of Nuclear Physics, 630090 Novosibirsk, 
  Russia}\thanks{email:telnov@inp.nsk.su}} 
\begin{document}
\newcommand{\EP}{\mbox{e$^+$}}
\newcommand{\EM}{\mbox{e$^-$}}
\newcommand{\EPEM}{\mbox{e$^+$e$^-$}}
\newcommand{\EMEM}{\mbox{e$^-$e$^-$}}
\newcommand{\GG}{\mbox{$\gamma\gamma$}}
\newcommand{\GE}{\mbox{$\gamma$e}}
\newcommand{\TEV}{\mbox{TeV}}
\newcommand{\GEV}{\mbox{GeV}}
\newcommand{\LGG}{\mbox{$L_{\gamma\gamma}$}}
\newcommand{\EV}{\mbox{eV}}
\newcommand{\CM}{\mbox{cm}}
\newcommand{\MM}{\mbox{mm}}
\newcommand{\NM}{\mbox{nm}}
\newcommand{\MKM}{\mbox{$\mu$m}}
\newcommand{\SEC}{\mbox{s}}
\newcommand{\CMS}{\mbox{cm$^{-2}$s$^{-1}$}}
\newcommand{\MRAD}{\mbox{mrad}}
\newcommand{\IND}{\hspace*{\parindent}}
\newcommand{\E}{\mbox{$\epsilon$}}
\newcommand{\EN}{\mbox{$\epsilon_n$}}
\newcommand{\EI}{\mbox{$\epsilon_i$}}
\newcommand{\ENI}{\mbox{$\epsilon_{ni}$}}
\newcommand{\ENX}{\mbox{$\epsilon_{nx}$}}
\newcommand{\ENY}{\mbox{$\epsilon_{ny}$}}
\newcommand{\EX}{\mbox{$\epsilon_x$}}
\newcommand{\EY}{\mbox{$\epsilon_y$}}
\newcommand{\BI}{\mbox{$\beta_i$}}
\newcommand{\BX}{\mbox{$\beta_x$}}
\newcommand{\BY}{\mbox{$\beta_y$}}
\newcommand{\SX}{\mbox{$\sigma_x$}}
\newcommand{\SY}{\mbox{$\sigma_y$}}
\newcommand{\SZ}{\mbox{$\sigma_z$}}
\newcommand{\SI}{\mbox{$\sigma_i$}}
\newcommand{\SIP}{\mbox{$\sigma_i^{\prime}$}}
\date{}
\maketitle
\begin{abstract}
 
  The idea of  plasma suppression of beam-beam effects at muon
  colliders is discussed. It is shown that one should take into
  account collisions in the plasma that were ignored before. Rough
  estimates show that this effect leads to a fast ``recovery'' of
  the beam magnetic field. For  beam parameters characteristic for
  muon colliders the suppression of the magnetic component of the beam
  field (1/2 of the total force) is almost absent.  It is also shown
  that the presence of the dense plasma (Li jet) at the interaction
  point leads to enormous hadronic background (due to photo-nuclear
  reactions) in the detector, about $10^{7}$ particles per crossing at
  large angles which creates serious problems for experimentation.

\end{abstract}

\section{Introduction}

One of main the problems for high energy muon colliders is the
limitation of the luminosity due to beam-beam interactions. A measure
of the beam-beam interaction is the tune-shift parameter
$\xi$~\cite{Wid}. For round beams, $\xi= Nr_c/4\pi\epsilon_n$, where
$r_c=e^2/mc^2$ is the classical radius of the beam particles, $\EN\ $
is the normalized transverse emittance. The value of $\xi$ should be
small enough ($\xi_{max}<0.1$) - otherwise the beams are disrupted due
to resonance diffusion. The maximum luminosity $L_{max} =N\gamma
f\xi_{max}/r_c \beta$, where $\beta$ is the $\beta$-function (usually
$\beta \approx \sigma_z$), $f$ is the collision rate. So, $L \propto
\xi$. This effect puts a severe limit on the luminosity of muon
colliders.
  
  One of the possible solution of this problem is  plasma suppression of
beam-beam interactions~\cite{Whittum,Stupakov,Lotov}. In a sufficiently
dense plasma one can expect that the induced charges and currents will
decrease the beam fields and, consequently, the  beam-beam effects. If plasma
decreases the beam field by a factor K, one can use beams with K times smaller
$\epsilon_n$ (to keep $\xi=\xi_{max}$), correspondingly the luminosity will be
K times larger.

It is essential to reduce both electric and magnetic fields because in
the vacuum their action on the opposing beam are equal in  value
and direction, the effective beam field is $|E|+|B|$. If the plasma
density $n_p$ is larger than the particle density of the colliding
bunches $n_b$, the electric field of the beam will be suppressed by
repelling (in the case of negatively charged bunch) or attracting (in
the case of positively charge bunch) plasma electrons (ions are
immobile). The nature of the magnetic field suppression is somewhat more
complicated. In the linear approximation the resulting suppression of
the beam-beam interaction~\cite{Stupakov}
\begin{equation}
\xi/\xi_0 \sim 4/(k_p\sigma_r)^2  \;\; \mbox{for} \;\;k_p\sigma_r >> 1,
\end{equation}
where $k_p=\omega_p/c$, $\omega_p^2=4\pi n_pe^2/m_e$, $\sigma_r$ is the
r.m.s. beam radius. A more accurate result, including nonlinear effects
and finite plasma thickness, was obtained in ref.~\cite{Lotov}.
   
So, for a factor of 5 suppression of beam-beam interactions the plasma
should satisfy the following requirements: a) $n_p>n_b$ and b) $k_p
\sigma_r > 4$. For example, let us take parameters of the
``evolutionary'' 100 TeV muon collider (see the B.King's table) but
with 5 times smaller $\ENX$: $N=0.8\times 10^{12}$,
$\sigma_r=\sqrt{2}\sigma_x=0.13\;\MKM,$ $\sigma_z=0.25$ cm. In this
case $\xi=0.5$ without suppresion, while the acceptable $\xi=0.1$. The
plasma density required for decreasing $\xi$ by a factor of 5 is found
from conditions a) and b), which give $n_p > 2.3\times 10^{21}$ and
$n_p > 2.5\times 10^{22}$, respectively.  As a source of plasma one
can use a liquid Li jet~\cite{Lotov} with electron density $1.5\times
10^{23}\;$ cm$^{-3}$. Such a target will be fully ionized by the muon
beam and the return current. If this theory is correct, in the
considered example one can increase the luminosity by a factor of 5.
With other beam parameters one can expect even better results, up to
30 for the given beam diameter.  All this sounds nice. However, there
are two effects which create serious problems for this method, and,
perhaps, close it:
\begin{itemize}
\item collisions in the plasma; 
\item hadronic background at large angles.
\end{itemize}
   
\section{Collisions in the plasma.}

In all papers on plasma suppression of the beam-beam interaction it was
assumed that plasma is collisionless. This picture is not correct. In  fully
ionized plasma the electrons of the return current do not lose energy
on  ionization, do not lose energy in collisions with other
electrons, because all electrons move with the same average velocity,
and have  very small energy loss in collisions with the ions; however, their
longitudinal velocity is decreased due to the scattering on the ions.

 The change of the longitudinal velocity in one scattering on the ion
\begin{equation}
\Delta v = -2v_0 \sin^2{\frac{\vartheta}{2}} \approx -v_0 \frac{\vartheta^2}{2}
\end{equation}
The resulting friction in the plasma
$$
\frac{ d\vec{v}}{dt} = -\int v_0\Delta v n_p d\sigma = -\int v_0^2
\frac{\vartheta^2}{2}8\pi n_p \left(\frac{e^2Z}{m_e v_0^2}\right)^2
\frac{d\vartheta}{\vartheta^3} = 
$$
\begin{equation}
= -4\pi n_p
\frac{\vec{v}}{v_0}\left(\frac{e^2Z}{m_e v_0}\right)^2
\ln{\frac{\vartheta_{max}}{\vartheta_{min}}} =
-4\pi
n_p \frac{\vec{v}}{v_0} \left(\frac{e^2Z}{m_e v_0}\right)^2 \ln{\Lambda},
\label{dv}
\end{equation}
where $\ln{\Lambda} = \ln({b_{max}/b_{min}})$. The minimum value
of the impact parameter $b$ follows from the energy conservation
$b_{min}=e^2Z/mv_0^2$, and the maximum value of $b$ is equal to the Debey
length $\lambda_D = \sqrt{kT/m}/\omega_p$. For the considered plasma densities
and electron velocities, $\ln{\Lambda} \approx 7 \sim 10$.

For estimation of the collision time one can take $dv = v$, which gives
\begin{equation}
\tau_{col} \sim \frac{v_0}{4\pi
n_p  \left(\frac{e^2Z}{m_e v_0}\right)^2 \ln{\Lambda}}.
\label{tau1}
\end{equation} 
The average velocity of electron in the return current $u\approx (n_b/n_p)c$. 
Although this velocity is not the same as $v_0$ due to transverse motion, 
let us  the first assume  $v_0=u$. Then
\begin{equation}
\tau_{col} \sim  (\frac{n_b}{n_p})^3   \frac{1}{4\pi c
n_p  Z^2 r_e^2 \ln{\Lambda}}.
\label{tau2}
\end{equation}
For the example given above $n_b/n_p\sim 1/60$, $Z=3$, $n_p=1.5\times
10^{23}$, that gives $\tau_{col} \sim 10^{-17}$ sec, which is much
smaller than the bunch collision time $\sigma_z/c\sim
0.25/3\times10^{10} \sim 10^{-11}$ sec. 

So, the assumption of collisionless plasma is not valid. The plasma
should be considered as a medium with some conductivity
$\sigma_c$. Accurate calculation of conductivity is a complicated
task because the electron drift in the longitudinal induction electric
field with very small energy loss, only scatter. Due to the field
their total kinetic energy continuosly grows, while the drift velocity
is approximately constant: $u\sim c(n_b/n_p)$. Nevertheless, we can make
some estimate.

  Eq.\ref{dv} is approximately valid even in the case of the ``hot''
return current if we will consider $\vec{v}$ as the drift velocity (from now
on $u$) and $v_0$ as the ``thermal'' velocity, which is still
unknown. The loss of the drift velocity given by Eq.\ref{dv} is
compensated by the induction electric field
\begin{equation}
|du/dt| = eE_{||}/m_e.
\label{c1}
\end{equation}
The conductivity is defined by equation
\begin{equation}
en_pu = \sigma_cE.
\label{c2}
\end{equation}
 Drift velocity is known:
\begin{equation}
u = c(n_b/n_p).
\label{c3}
\end{equation}
Using Eqs~\ref{c1},\ref{c2},\ref{c3}, we get
\begin{equation}
\sigma_c =\frac{e^2 n_b c}{m_e|du/dt|}
\label{c4}
\end{equation}
Using Eq.\ref{dv}, we obtain
\begin{equation}
\sigma_c =\frac{m_e v_0^3}{4\pi e^2 Z^2 \ln{\Lambda}}
\label{c5}
\end{equation}

Now we have to estimate $v_0$. The collision time is given by
 eq.\ref{tau1}.  Between two collisions the electron drift velocity is
 restored by the induction electric field and the total energy is
 increased by about $m_eu^2$, so as an estimate one can take the
 average kinetic energy to be equal to $m_ev_0^2 \sim N_{col} m_e
 u^2$.  The maximum number of scatterings for the same electron can be
 estimated as the number of collisions after which its transverse
 displacement is equal to the beam radius (then this electron in the
 return current is replaced by the new one which comes from outside
 and is initially cool)
\begin{equation}
\tau_{col} v_0\sqrt{N} \sim \sigma_r.
\label{c61}
\end{equation}
Using this arguments and Eq.\ref{c5} we find 
\begin{equation}
v_0 = c (4\pi n_b Z^2 \sigma_r r_e^2 \ln \Lambda )^{1/5}
\label{c7}
\end{equation}
For Z=3 (Li), $n_b =2.5\times 10^{21}$ (see the example above) and
$\sigma_r= 0.15$ \MKM, $v_0 \sim 0.075$c.

So, the conductivity is found, see Eqs~\ref{c5},\ref{c7}. Now we have to
understand how the conductivity influences  the plasma suppression of the
bunch field.  The return current is driven by the longitudinal
electric field $E_{||}$ that is caused by  penetration of the beam
magnetic field into the plasma. From Faraday law
\begin{equation}
E_{||} \sim \frac{1}{c}\frac{d(B_{\phi}\sigma_r)}{dt}.
\label{c8}
\end{equation}
This electric field produces the  return current equal approximately  to the
bunch current $I$ (if compensation works)
\begin{equation}
\sigma_c E_{||} \pi \sigma_r^2 \sim I. 
\label{c9}
\end{equation}
Introdusing $B_{\phi} \sim 2I/c\sigma_r$ (the beam field when there is no
beam field suppression) and Eqs~\ref{c8},\ref{c9}, we obtain
\begin{equation}
\frac{dB_{\phi}}{B_{\phi}} \sim \frac{c^2 dt}{2\pi \sigma_r^2 \sigma_c}.
\label{c10}
\end{equation} 
The relative value of the beam field which penetrates into the plasma
during the time of the bunch collision is
\begin{equation}
\frac{\Delta B}{B} \sim \frac{c \sigma_z}{2\pi \sigma_r^2 \sigma_c} \sim
\frac{2\sigma_z r_e Z^2 \ln{\Lambda}}{\sigma_r^2(v_0/c)^3},
\label{c11}
\end{equation} 
where $v_0$ is given by Eq.\ref{c7}. For the example considered in
this paper: $\sigma_z=0.25$ cm, $Z=3$, $\sigma_r =0.13$ \MKM,
 $N=0.8\times10^{12}$, $v_0/c \sim 0.075$, $\ln{\Lambda} =7$ we get 
\begin{equation}
\frac{\Delta B}{B} \sim 100\;!!!
\label{c12}
\end{equation} 
In order to obtain plasma suppression by one order of magnitude we need
$\Delta B/B \sim 0.1$. So, it seems that plasma does not
help.  Although my estimate is very approximate, it is very unlikely that
a factor of 1000 is lost. 

\section{Backgrounds}
\subsection{Photo-nuclear reactions}

The total photo-nuclear cross section for lithium is~\cite{partdata}
 \begin{equation}
\sigma_{\gamma Li} \sim 0.4\times 10^{-27}\;\mbox{cm}^2.
\end{equation} 
The number of virtual photon with the energy above 1 GeV per one muon
 at 100 TeV muon collider is
\begin{equation} N_{\gamma} \sim \int
 \frac{2\alpha}{\pi} \ln \left(\frac{E}{\omega}\right)\frac{d\omega}{\omega} 
 \sim 0.3 N_{\mu}.
\end{equation} 
The number of  ph.n. reactions per bunch crossing generated by
$2\cdot 10^{12}$ muons (two beams) in  $l=0.5$ cm Li jet is
\begin{equation}
N_{b} = N_{\gamma}n_{Li} l \sigma_{\gamma} \sim
0.3\times 2\cdot 10^{12} \times 5\cdot 10^{22} \times 0.5 \times
0.4\cdot 10^{-27} = 0.6\cdot 10^7\;!
\end{equation} 
Although most of the produced particles travel in the forward
direction, each reaction produce gives approximately one particle
($\pi^{\pm},\pi^0$) at large angles, with $P \sim P_t \sim 300$
MeV. The total energy of these particles is greater than $ 2\cdot 10^3$ TeV.

    It is hard to imagine a detector which could work in
conditions so terrible!
\subsection{\EPEM\ production: $\mu Li \to \mu Li \EPEM$}

The cross section of this reaction~\cite{Budnev} 
\begin{equation}
\sigma \approx \frac{28\alpha^2 r_e^2}{27\pi}(l^3-6.36l^2)(Z_1Z_2)^2,
\end{equation}
where $l=\ln \frac{2(P_1P_2)}{m_1m_2} \approx \ln {2\gamma_{\mu}}$. In
the case of 50 TeV muons and the Li target, $\sigma = 1.8\times
10^{-26}$ cm$^2$

The probability of \EPEM\ pair creation by a muon in a 1 cm thick Li
jet for 1000 crossings (as it is in the muon colliders) is about
$1-e^{-1}$ (one interaction length). In most cases the energy loss
is not large but sufficient to knock  the muon out of the $10^{-4}$
energy range which contributes to  luminosity (muons with larger
energy deviations are defocussed due to chromatic abberations of the
final focus system).  So, this effect will decrease the luminosity
lifetime by about a factor of 2.

\section{Conclusions}

   Suppression of the beam-beam effects by a dense plasma jet (Li) at
the collision point is a very attractive idea. However, collisions in
the plasma significantly change the picture. This effect was ignored
before. Rough estimates show that this effect leads to fast
``recovery'' of the magnetic beam field, leaving it practically
unsuppressed. This result should be checked by more accurate
calculations.

  Photo-nuclear reactions produce enormous hadronic backgrounds in the
detector ($\sim 10^7$ particles/crossing at large angles), so the
possibility of experimentation at such background conditions is
practically impossible.  Electro-production of \EPEM\ pairs in Li jet
leads to some decrease of the luminosity lifetime.

\vspace{1cm}
  I would like to thank K.Lotov and A.Skrinsky for useful discussions
and B.King for organization of the very fruitful workshop.

\end{document}